\documentclass[conference]{IEEEtran}
\usepackage[margin=1in]{geometry}
\usepackage{blindtext, graphicx}
\usepackage{amsmath}
\usepackage{wrapfig}
\usepackage{graphicx}
\usepackage[english]{babel}
\usepackage{blindtext}
\usepackage{draftwatermark}

\SetWatermarkText{PREPRINT}
\SetWatermarkScale{1}

\begin{document}

\title{A multinomial probabilistic model for movie genre predictions}

\author{\IEEEauthorblockN{Eric Makita, Artem Lenskiy}

\IEEEauthorblockA{Korea University of Technology and Education\\
1600, Chungjeol-ro, Byeongcheon-myeon,
Dongnam-gu, Cheonan-si, Chungcheongnam-do 31253,\\ Republic of Korea\\
email: ericmakita@koreatech.ac.kr, lensky@koreatech.ac.kr}
}

\maketitle

\begin{abstract}

This paper proposes a movie genre-prediction based on multinomial probability model. To the best of our knowledge, this problem has not been addressed yet in the field of recommender system. The prediction of a movie’s genre has many practical applications including complementing the item’s categories given by experts and providing a surprise effect in the recommendations given to a user. We employ mulitnomial event model to estimate a likelihood of a movie given genre and the Bayes rule to evaluate the posterior probability of a genre given a movie. Experiments with the MovieLens dataset validate our approach. We achieved 70\% prediction rate using only 15\% of the whole set for training.

\end{abstract}


\IEEEpeerreviewmaketitle

\section{Introduction}
Nowadays web users are confronted with an overabundance of information caused by a constant increase in the volume of information that users can no longer absorb, process, or prioritize. As a result, the selection and use of information processed by users become particularly complex. This phenomenon presents the need to design a recommender system that best depicts the preferences of a user with regard to the most relevant information at possibly the shortest time \cite{ref1,ref2}.\newline
\indent Recommender systems are tools whose purpose is to help users overcome information overload by selecting the most interesting information based on their preferences. In other words, they try to predict a user's interest towards an item. The items to recommend are varied, ranging from movies to watch, books to read, podcast to listen to, or else. Conventional recommender systems have been successfully applied by e-commerce or social networking websites such as Amazon (www.amazon.com), Netflix (www.netflix.com), YouTube (www.youtube.com), and Facebook (www.facebook.com). Since then many recommender system algorithms and their variants have been proposed in literature, however, most of them were mainly accuracy-oriented algorithms that predict the rating of an item. In other words, these algorithms were focused on optimizing the accuracy of the rating of a predicted item. Although such recommender system algorithms are sufficient in many applications, there are situations where they may not be enough because they do not take into account all facets of the user's interests such as the desire to change. At this point the need to design a new paradigm of recommender systems arises considering important factors outside the optimization of the accuracy of the rating of predicted item. \\
\indent A multinomial probabilistic model for movie genre prediction is proposed in this paper as a response to the confronting challenge. Our model aims to predict a movie's genre rather than to predict its rating. Technically, this approach is based on Bayesian reasoning and is enacted in two steps. In the first step, the multinomial probabilistic model is applied to learn the movie's likelihood of belonging to a particular genre using multinomial model. At the second step, the Bayesian probabilistic reasoning is applied for the prediction of the movie's genres. This in turn completes the item's genre given by experts and consequently can improve the recommendation.\\
In the recent past, many recommender systems that consider factors other than the accuracy-oriented ones and their variants have been proposed in literature \cite{ref3,ref4,ref5,ref6}. To the best of our knowledge, our model predicting an item's genre/s that complement the genres assigned by experts is a new attempt in the field of recommender systems. The rest of the paper is organized as follows. Section 2 discusses the related work. Section 3 outlines the proposed algorithm. Section 4 contains the performance study. Finally in section 5, we summarize our work.

\section{Related work}

Traditional recommender systems focus on increasing the accuracy of ratings prediction of items not yet viewed by active users based on ratings of items already viewed by active users. However, in \cite{ref7},  McNee  et  al. stated that focusing only on improving the accuracy of the recommended items' rating is not enough for improving the user satisfaction. Factors such as coverage, diversity, novelty and/or serendipity should be used to improve the recommendations. Aiming at enhancing the conventional recommender systems while working with important factors outside  item  rating  prediction  accuracy,  many studies developing new  paradigm of making recommendation were proposed in the literature. We briefly discuss related works considering fcators outside of accuracy  in recommender systems and survey works that have done new attempts in the area of recommender systems.  \\
\indent Saul et al. \cite{ref8} enhanced the diversity of recommendations by adopting the re-ranking approach with a greedy selection. This approach allows to discover new useful recommendations that were not searched by the user. In \cite{ref9}, Tuzhilin and Adamopoulos proposed a probabilistic neighborhood selection in collaborative filtering that directly generates diversified recommendation lists prior to overcoming the overspecialization issues in traditional recommendations. Recent researches have been focused on designing new attempts in the literature. Liu et al. \cite{ref10} proposed a novel approach of collaborative filtering whose goal is to select a small set of special users called star users. Unlike traditional approaches, collaborative filtering allows finding similar users or neighborhood for each target user. Collaborative filtering based on star users employs star users to represent the interest of the whole set of users prior to computing the recommendation, so as to improve the scalability of the traditional recommender systems. \\
\indent Amatriain et al. \cite{ref11} proposed a variation of the traditional collaborative filtering approach that uses expert opinions from independent data set to compute the recommendations. In this approach, the nearest neighbor-oriented recommendation was scaled-up by the use of these external experts.

\section{The Naive Bayes Classifier for text classification}

In order to solve classification problems, many methods such as linear discriminant analysis, neural networks, support vector machines, etc have been proposed. Among them, one of the oldest and well-known classification algorithms is the Naive Bayes classification algorithm which dates back to 18th century. The algorithm performs well in various tasks despite its simplicity, however it comes with a naive assumption that features that constitutes features vectors are independent. One of the well-known applications of Naive Bayes is the text classification. The details of the Naive Bayes classifier are well summarized in \cite{ref12,ref13}. This section will provide a brief overview of the Naive Bayes classifier applied to text classification.\\
\indent The Naive Bayes classifiers is based on the Bayesian theorem and is appropriate for multidimensional data i.e. multidimensional feature vectors. Regarding the text document classification problem, a document is represented as a bag of words where each word is part of a vocabulary $V$, $ w \in V $. Considering each document $ d \in D $ where $ D $ represents the set of all training documents and every document is labeled as $c$ where $c \in C $ is a set of $\lvert C \rvert$ distinct classes. According to the Bayes theorem the Naive Bayes classifiers estimates the probability of a class $c$ given a document $ d $ as follows:
\begin{equation}
P( c \mid d ) = \frac{P(c) \cdot P( d\mid c )}{P(d)}
\label{eq:1}
\end{equation}
\indent This estimate can be further used for document classification by simply assigning the class label that corresponds to the highest probability $P( c \mid d )$. The probability $P(c)$ is the prior probability of a document being of class $c$ without actually knowing anything about the document.  The prior class probability $P(c)$ is estimated by counting the number of training  documents in every class: \\
\begin{equation}
P(c)=\frac{\sum_{d \in D}P(   c \mid d)}{\lvert D \rvert}
\label{eq:2}
\end{equation}
where $P(c\mid d) \in \{0, 1\}$, is 1 if document is labeled as $c$ otherwise it is 0 and $\lvert D \rvert$  is the total number of training documents.
The conditional probability  $P( d\mid c )$ incorporates information about the document via the set of words that occurred in the document. Depending on a class $c$ of a document, the probability of a word occurrence $P( w\mid c )$ will vary as well as $P(d \mid c )$. To simplify the computation of the $P(d \mid c )$, $d = \{w_{1},w_{2},...,w_{\lvert d \rvert}\}$, where  $\lvert d \rvert$ is the  document  length,  it is assumed that words are conditionally independent, i.e. the occurrence of words do not depend upon each other.  Thus, $P(d \mid c) = P( w_{1}\mid c ) \cdot P( w_{2}\mid c )...P( w_{\mid d \mid }\mid c )$. A slightly different form of computing  $P(d \mid c)$ is given as follows
\begin{equation}
P(d \mid c) =  \prod\limits_{w \in V}P( w\mid c )^{N(w)}
\label{eq:3}
\end{equation}
where $w_{i}$ is a unique word in the vocabulary $V$. Here, instead of iterating over the words in a document we iterate over the set of all words in the vocabulary $V$. Notice that some words from the vocabulary might occur multiple times in that case $1 < N(w)$, and on the other hand some words might not occur in the document at all, in such a case $N(w) = 0$. Substituting eq. \ref{eq:3} to eq. \ref{eq:1} and assuming  $P(d_i) = P(d_j), \forall i,j$, i.e. all documents are equally to occur, the classification can be performed as follows :
\begin{equation}
c = \arg\max_{c} P(c)\prod\limits_{w \in V}P( w\mid c )^{N(w)}
\label{eq:4}
\end{equation}

Probabilities $P( w_\mid c )$ of a word $w$ occurring in a class $c$ is estimated by counting the number of times a word $w$ appeared in all documents of class $c$. However, some words might not appear at all in documents of a particular class resulting in a zero probability. To avoid the zero probability in such cases a small probability is assigned according the Laplace's Law of succession:
\begin{equation}
P( w\mid c)=\frac{1 + N(w,c)}{\lvert V \rvert + \sum_{w \in V} N(w, c)}
\label{eq:5}
\end{equation}
 where $N(w,c) $ represents the number of times the word $w$ occurred in the training documents whose class label is $c$, $N(c)$ denotes the number of documents labeled as $c$. \\


\section{Proposed method}

We propose to employ a similar approach of the text-classification\cite{ref14} discussed in the previous section to predicting movies' genres. Following the approach of bag-of-words for text classification we propose a bag-of-users approach to predict movies' genres. Such prediction is possible due to the fact that users are usually consistent with their preferences, and prefer some genres over others.\newline
\indent In the following, we introduce symbols and definitions used in this section. Let $U$ be the set of all users, $ M $ be the set of all movies and $ G $ be the set of genres. The movie $ m \in M $ is represented by $\lvert U \rvert$ dimensional feature vector $ r_m = \{\mathbf{R}_{u_1,m}, \mathbf{R}_{u_2,m},...,\mathbf{R}_{u_{\vert U \rvert}, m}\}$, where $\mathbf{R}$ is rating matrix and $\mathbf{R}_{u_i,m}$ is either a 0 or a 1, depending if the user $u_i$ rated a movie $m$ or not.\\
\indent It is assumed that users are conditionally independent. That is, the ratings of one user do not depend on ratings of another user. This assumption is required by the Naive Bayes classifier. Thus, the movie likelihood $P( m \mid g )$, can be written as a categorical distribution, that is a particular case of multinomial distribution:

\begin{equation}
P( m \mid g ) \propto \prod\limits_{u \in U}P(u \mid g)^{\mathbf{R}_{u, m}}
\label{eq:6}
\end{equation}

where $\mathbf{R}_{u, m}$ represents whether the user $u$ rated a movie $m$ or not.  We use categorical distribution because a user does not rate the same movie twice, whereas in the text classification the same word can appear in the text more than once. It is often the case that ratings are not binary. There could be more than one rating. For example, 0 corresponds to no rating, 1 to a dislike, 2 to a neutral response and 3 to a like. Then matrix $\mathbf{R}$ is split into three binary matrices $\mathbf{R}_1$, $\mathbf{R}_2$, $\mathbf{R}_3$.  Then an estimate of the probability $P(u \mid g )$ is computed as follow:

\begin{equation}
P( u \mid g ) =\frac{1+ N(u,g)}{\lvert|U \rvert+\sum_{u \in U}N(u,g)}
\label{eq:7}
\end{equation}

This probability is estimated as a relative frequency of user's $ u $ ratings of movies labeled as genre $ g $ to the total number of users' ratings of the movies from that genre. \newline 
\indent Note that the Laplace’s Law of succession was used to avoid the zero probability estimates that can occur for situations where a user $ u $ did not rate a movie $m$.

Giving a movie training set $ M $,  the prior probability of $g$ can be estimated as the relative frequency of movies of genre $g$ to the total number of movies:

\begin{equation}
P(g)=\frac{\sum_{m \in M}P(g \mid m)}{\lvert M \rvert}
\label{eq:8}
\end{equation}
where $ P(g \mid m) $ takes the value of $1$ if $m$ is marked as genre $g$, $0$ otherwise. In the case when a number of $N$ genres are assigned to a movie, the probability is corrected as follows $ P(g \mid m) = 1/N$.    \newline

Figure 1 shows the preferences of all of the users in the training set calculated using eq. \ref{eq:7}. Figure 2 illustrates  the conditional probability calculated according to eq. \ref{eq:6} with prior taking the logarithm. 

\begin{figure}[h]
\centering
\includegraphics[trim=3.5cm 8cm 4cm 7.5cm, width=0.4\textwidth]{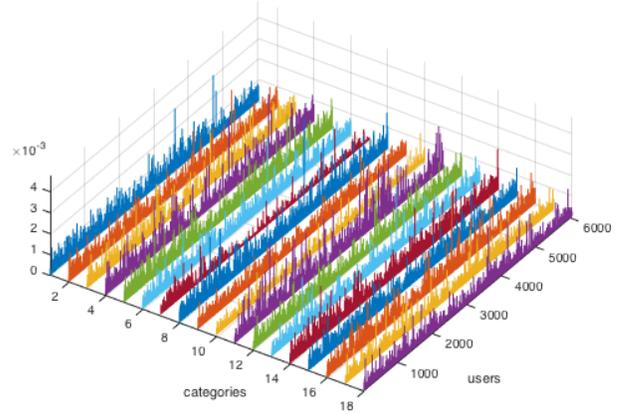}
\caption{The user preferences represented by $P( u \mid g )$.}
\end{figure}

\begin{figure}[h]
\centering
\includegraphics[trim=3.5cm 8cm 4cm 7.5cm, width=0.4\textwidth]{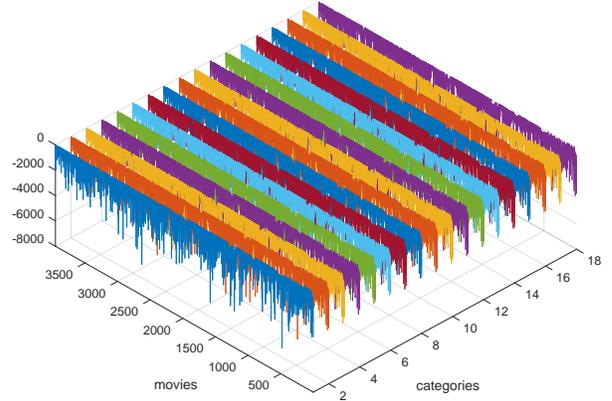}
\caption{The logarithm of conditional probabilities $P( m \mid g )$.}
\end{figure}

\section{Experimental evaluation and discussion}

To evaluate the effectiveness of the proposed methodology of predicting movie genres using users’ ratings, we employed the 1M Movielens dataset. The dataset contains one million ratings of 3952 movies given by 6040 users. Each user rated at least 20 movies on the scale from 1 to 5, i.e. dislike, slightly dislike, neutral, slightly like and like. A rating of 1 shows how strongly a user dislikes a movie while a rating of 5 shows how strongly a user likes a movie. The experiments were carried out in the following procedure: a portion of the movies was randomly selected for training and the remaining movies were used to assess the quality of genre prediction. The portion of the movies used for training varied from 5\% to 95\% with a 5\% step, so overall we experimented with 19 different training sizes. For every portion we repeated the experiment 50 times, then estimated the mean and the standard deviation. Among all movies about 33\% were assigned two genres, 10\% three genres, and less than 3\% four, five and six genres, and the remaining 51\% of the movies were labeled with only one genre. We considered the prediction is successful if the predicted genre matches the true one, or one of the true ones if a movie simultaneously was assigned to a number of genres. Prediction was performed separately for every rating. Figure 3 summarizes the evolution of genre prediction.

\begin{figure}[h]
\centering
\includegraphics[trim=4.5cm 10cm 5cm 9cm, width=0.4\textwidth]{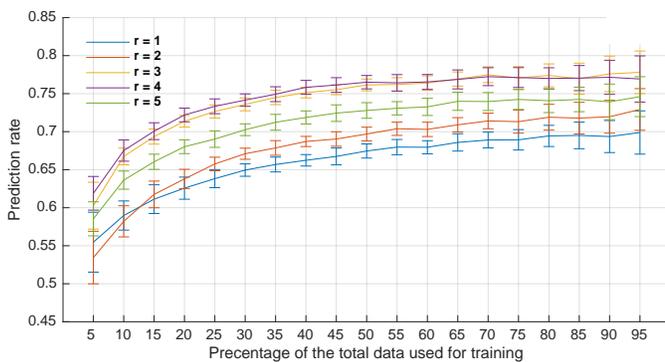}
\caption{Genres' prediction rate depending on the size of the training set.}
\end{figure}

As it can be seen in the figures, the more movies are used in the training the better prediction is. However, an interesting observation here is that even 5\% of the data used for training is enough to obtain 62\% prediction rate. If the prediction was performed simply by using the prior probability, only 30\% of the movies would be predicted correctly. Increasing the training set does improve the prediction rate however, the improvement is no as significant as for the first 5\%. For instance, increasing the size of the training set to 95\% would only improve the prediction rate by a slightly more than 15\%.  From this analysis we are ought to conclude that even a small portion of information about the users' ratings is enough to learn users’ preferences that allows us to predict genres. Particularly, the multinomial model was capable of predicting movies' genres based on users' preference profiles well enough. 
The second interesting result is that better predictions are achieved for neutral responses i.e. movies with ratings 3 and for slightly likes i.e. movies with ratings 4, rather than for the expected rating 5. The worst predictions are based on the rating 1 and 2. These results might be ambiguous, however after counting the number of times ratings were given in every rating category, we found that the most popular ratings are 3 and 4 (fig. 4). Thus, the higher number of ratings provides higher prediction accuracy. 
Currently, the prediction was performed separately for every class of ratings, for instance, only based on strong dislikes. As for the further research we are planning to investigate if some combinations of ratings will improve the prediction.

\begin{figure}[h]
\centering
\includegraphics[trim=4.5cm 10cm 5cm 9cm, width=0.4\textwidth]{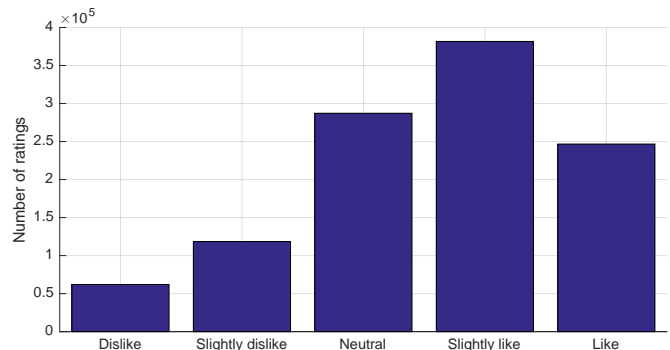}
\caption{Number of ratings of each type given in the whole set.}
\end{figure}

\section{Conclusion and future work}

Recommender systems are tools whose purpose is to help users overcome information overload by selecting the most interesting information based on users' preferences and also to help companies sell more products. In this paper, we proposed to predict a movie’s genre rather than predicting movie’s ratings. For this purpose, we applied Naïve Bayes classifier and multinomial event model, that previously was used for text classification and a number of other tasks. This approach can potentially improve the efficiency of recommendation by recommending movies from unexpected but relevant genres, this will be investigated in further research. We showed that the genre prediction rate increases while the size of the training data increases, however the rate of prediction slows down. Even a small number of users allows significantly increase genre prediction. That is, overall, we successfully applied ratings for genre prediction in the context of movies.

\ifCLASSOPTIONcaptionsoff
  \newpage
\fi

\vspace{-120 mm}
\begin{IEEEbiography}
    [{\includegraphics[width=1in,height=1.25in,clip,keepaspectratio]{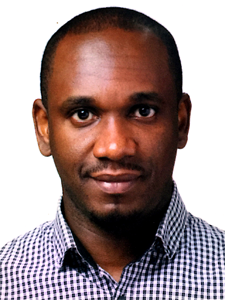}}]{Eric Arnaud Makita Makita} holds BSc and MSc degrees in computer science and engineering from the Institut Supérieur d’Informatique and Ecole Supérieure de Technologie et de Management in Dakar, Senegal in 2005 and 2007, respectively. Currently he is a Ph.D. candidate at the Korea University of Technology and Education.
\end{IEEEbiography}
\vspace{-120 mm}
\begin{IEEEbiography}
    [{\includegraphics[width=1in,height=1.25in,clip,keepaspectratio]{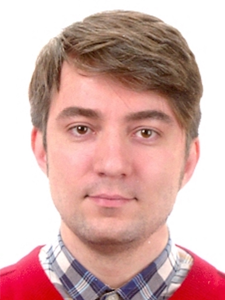}}]{Artem Lenskiy} Received BSc and MSc degrees in computer science from Novosibirsk State Technical University, Russian in 2002 and 2004, respectively.He joined doctor course at the University of Ulsan, Korea. He was awarded the Ph.D in 2010 from the same university. After conducting research as a postdoc fellow at the Ulsan University, he joined Korea University of Technology and Education as an assistant professor in 2011.\\
    His research interests include machine learning and self-similar processes applied to various research and engineering fields including analysis of financial time series, telecommunication and physiological signals.
\end{IEEEbiography}
\end{document}